\documentclass[sigconf,natbib=true]{acmart}
\usepackage{subfig}
\usepackage{hyperref}
\usepackage{multirow}
\usepackage{xspace}
\usepackage{enumitem}
\newcommand{\modelname}{\textsf{STARCRS}\xspace} %

\AtBeginDocument{%
  }

\acmISBN{978-1-4503-XXXX-X/2018/06}

\begin{document}

\title{Integrating Vision-Centric Text Understanding for Conversational Recommender Systems}

\author{Wei Yuan}
\affiliation{%
  \institution{The University of Queensland}
  \city{Brisbane}
  \state{QLD}
  \country{Australia}
}
\email{w.yuan@uq.edu.au}

\author{Shutong Qiao}
\affiliation{
  \institution{The University of Queensland}
  \city{Brisbane}
  \state{QLD}
  \country{Australia}
}
\email{shutong.qiao@uq.edu.au}

\author{Tong Chen}
\affiliation{
  \institution{The University of Queensland}
  \city{Brisbane}
  \state{QLD}
  \country{Australia}
}
\email{tong.chen@uq.edu.au}

\author{Quoc Viet Hung Nguyen}
\affiliation{
    \institution{Griffith University}
    \city{Gold Coast}
    \state{QLD}
    \country{Australia}
}
\email{henry.nguyen@griffith.edu.au}

\author{Zi Huang}
\affiliation{
    \institution{The University of Queensland}
    \city{Brisbane}
    \state{QLD}
    \country{Australia}
}
\email{huang@itee.uq.edu.au}

\author{Hongzhi Yin}
\authornote{Corresponding author.}
\affiliation{
    \institution{The University of Queensland}
    \city{Brisbane}
    \country{Australia}
}
\email{h.yin1@uq.edu.au}

\renewcommand{\shortauthors}{Yuan et al.}

\begin{abstract}
  Conversational Recommender Systems (CRSs) have attracted growing attention for their ability to deliver personalized recommendations through natural language interactions. To more accurately infer user preferences from multi-turn conversations, recent works increasingly expand conversational context (e.g., by incorporating diverse entity information or retrieving related dialogues). While such context enrichment can assist preference modeling, it also introduces longer and more heterogeneous inputs, leading to practical issues such as input length constraints, text style inconsistency, and irrelevant textual noise, thereby raising the demand for stronger language understanding ability.
  In this paper, we propose \modelname, a \underline{s}creen-\underline{t}ext-\underline{a}wa\underline{r}e \underline{c}onversational \underline{r}ecommender \underline{s}ystem that integrates two complementary text understanding modes: (1) a screen-reading pathway that encodes auxiliary textual information as visual tokens, mimicking skim reading on a screen, and (2) an LLM-based textual pathway that focuses on a limited set of critical content for fine-grained reasoning. We design a knowledge-anchored fusion framework that combines contrastive alignment, cross-attention interaction, and adaptive gating to integrate the two modes for improved preference modeling and response generation. Extensive experiments on two widely used benchmarks demonstrate that \modelname consistently improves both recommendation accuracy and generated response quality. 
\end{abstract}



\maketitle

\section{Introduction}
Conventional recommender systems have achieved substantial success by leveraging users’ historical interaction signals to deliver personalized recommendations. However, they often struggle to accurately capture users’ real-time intentions and to provide effective recommendations for users with limited interaction histories, such as in cold-start scenarios~\cite{raza2026comprehensive}. To overcome these limitations, conversational recommender systems (CRSs) have been proposed. By engaging users in multi-turn natural language dialogues, CRSs can actively elicit richer preference signals and gain a deeper understanding of users’ immediate needs, thereby enabling more accurate, adaptive, and user-aligned recommendations~\cite{an2025beyond,zhu2025llm}.

Despite their promise, extracting accurate user preferences from free-form conversations remains challenging. Natural conversations typically contain sparse preference-related signals and a substantial amount of irrelevant or low-information content (e.g., chit-chat, acknowledgements, and digressions), making it difficult for models to identify what truly matters~\cite{zhang2025mitigating}.
To mitigate this issue, many CRSs incorporate external information to strengthen preference modeling. For instance, a line of work~\cite{chen2019towards,lin2023cola,zhou2020improving,zou2022improving} leverages knowledge graphs to inject entity information, highlighting salient entities mentioned in the dialogue and linking them to structured knowledge.
More recently, Dao et al.~\cite{dao2024broadening} proposed DCRS, which broadens each conversation by retrieving other users' dialogues as demonstrations. Building on this direction, MSCRS~\cite{wei2025mscrs} further uses LLMs to enrich entity information and adapt it into conversational contexts.

\begin{figure}[t]
    \centering
    \includegraphics[width=0.35\textwidth]{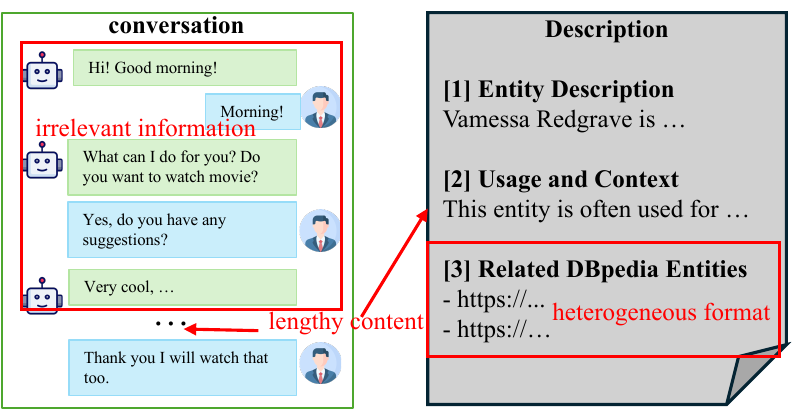}
    \caption{Problems in CRS's textual materials.}
    \label{fig_illustrate}
\end{figure}

While these approaches have improved both recommendation accuracy and response generation quality, they introduce a new challenge: CRSs are required to process larger volumes of text that are more diverse in form and often lower in quality. As illustrated in Figure~\ref{fig_illustrate}, the additional information incorporated in practice is frequently noisy, heterogeneous, and lengthy, substantially increasing the difficulty of effective language understanding. First, the incorporated content often contains a considerable amount of irrelevant information. Retrieved dialogue histories and LLM-augmented entity descriptions may include greetings, filler utterances, or overly verbose explanations that are only weakly related to users’ true preferences. Second, external information is typically presented in diverse formats. Some sources, such as retrieved dialogue contexts, are inherently sequential, whereas others, such as entity catalogs or attribute blocks, resemble semi-structured records with limited ordering constraints across fields. Treating all such inputs as a single flat token sequence can exacerbate format mismatches and hinder the model’s ability to identify preference-critical signals. Third, context expansion can easily push inputs toward the length limits of backbone language models. When truncation becomes unavoidable, crucial preference evidence may be discarded, while the model may over-attend to easily accessible but uninformative text. Consequently, even with rich auxiliary contexts, CRSs may still fail to accurately infer user preferences, resulting in suboptimal recommendation performance and low-quality conversational responses.

Recent advances in vision-centric text understanding offer a complementary perspective for processing complex and heterogeneous textual inputs~\cite{xing2025vision}. Rather than strictly linearizing text into a one-dimensional token sequence, these approaches render text as images and learn representations that bridge the semantic gap between visual layouts and their underlying linguistic content through carefully designed pretraining objectives~\cite{xing2025vision,rust2022language}. A representative example is DeepSeek-OCR~\cite{wei2025deepseek}, which projects textual tokens into a two-dimensional space that mirrors on-screen text layouts and pretrains a vision-centric encoder using OCR-related objectives to preserve semantic fidelity to the original text. Owing to their reduced reliance on strict token order and predefined vocabularies, vision-centric encoders can naturally accommodate layout-aware and block-structured content. As a result, they tend to be more robust to noise and better suited for encoding broader contexts~\cite{tai2024pixar}, making them particularly attractive for CRS context expansion.

Motivated by these advances, we take a first step toward incorporating vision-centric text processing into CRSs as an auxiliary text-understanding pathway, enabling multi-granularity modeling for preference extraction from complex contexts. We propose \modelname (\underline{s}creen-\underline{t}ext-\underline{a}wa\underline{r}e \underline{c}onversational \underline{r}ecommender \underline{s}ystem). The key idea is to render massive or relatively low-salience textual content into “screen text” and encode it with a pretrained vision-centric encoder for coarse-grained understanding, while still using a standard text encoder to process concise but critical textual evidence for fine-grained reasoning. This design mirrors human reading behavior: we often skimly read broaden content but carefully read and think about the most informative few sentences.

Concretely, for recommendation, we first prompt a pretrained LLM to generate comprehensive natural-language descriptions for entities. We then render the full descriptions as screenshot-style screen text and obtain coarse-grained, full-context embeddings with a pretrained vision-centric encoder, which is robust to long and heterogeneous content (e.g., extended descriptions, usage notes, and related-entity links). In parallel, we use a text encoder to process only a short, high-salience subset (e.g., the Entity Description part in Figure~\ref{fig_template}), truncated to fit the maximum input length, producing fine-grained semantic embeddings. \modelname aligns these complementary views with contrastive objectives and fuses them via knowledge-anchored cross-attention and adaptive gating to better capture user preferences for recommendations. For conversational response generation, we follow the same skim-read and careful-read idea. The text encoder ingests the conversation in its truncated textual form, typically retaining the most recent utterances under the length limit, while the vision-centric pathway encodes the entire dialogue history as screen text to provide a broader skim-reading signal. 

To sum up, the main contribution of the paper is threefold:
\begin{itemize}
  \item To the best of our knowledge, we are the first to improve conversational recommender systems through the lens of vision-centric text processing, a recently emerging paradigm that is particularly well suited for lengthy, noisy, and heterogeneous textual inputs.
  \item We propose \modelname, a dual-path design that encodes auxiliary “screen text” with a pretrained vision-centric encoder for coarse-grained context modeling, while leveraging a text encoder for fine-grained reasoning over concise, salient evidence. The two pathways are adaptively fused via contrastive learning and gated cross-attention to better capture user preferences and support response generation.
  \item We conduct extensive experiments on two widely used conversational recommendation benchmarks, together with thorough analyses, to verify the effectiveness of \modelname and each of its key components.
\end{itemize}

\section{Related Work}
\subsection{Conversational Recommender System}
Conversational recommender systems (CRSs) have been an active research direction because they can interactively elicit users' evolving preferences and provide recommendations in real time~\cite{jannach2021survey}. 
In general, CRSs can be grouped into two major paradigms according to the interaction form: attribute-based and generation-based CRSs.

Attribute-based CRSs aim to improve recommendation accuracy by collecting key user intents with as few interaction turns as possible~\cite{yu2025causal}. 
They typically rely on a predefined set of attributes and related questions, and select the most informative question based on the current dialogue state using techniques like reinforcement learning~\cite{deng2021unified,lei2020estimation,ren2021learning} or bandit-based strategies~\cite{li2021seamlessly,li2025towards}, thereby obtaining clarifications that reduce uncertainty and narrow down the candidate set. 
Once sufficient signals are gathered, the system produces a recommendation list for the user.

Generation-based CRSs, in contrast, target natural and flexible interactions in free-form language. 
They thus need to jointly optimize recommendation quality and the coherence, informativeness, and readability of generated responses. 
Early work improved performance mainly through tailored neural architectures and training strategies~\cite{li2018towards,chen2019towards}. 
With the rapid progress of large pre-trained language models (LLMs), recent CRSs increasingly adopt LLMs as the backbone and shift the emphasis toward context enrichment, augmenting dialogue history with external knowledge and evidence to better infer user preferences from otherwise sparse or ambiguous conversational cues.
For instance, Zhou et al.~\cite{zhou2020improving} leverage DBpedia~\cite{auer2007dbpedia} and ConceptNet~\cite{speer2017conceptnet} to highlight entities mentioned in conversations and enhance reasoning over user interests. 
Beyond knowledge graphs, other works incorporate additional textual evidence such as user reviews (e.g., RevCore~\cite{lu2021revcore}) to improve preference modeling. 
More recently, retrieval-augmented CRSs have been proposed to broaden dialogue context with similar users' conversations as demonstrations~\cite{dao2024broadening}. 
Wei et al.~\cite{wei2025mscrs} further extend the context with LLM-generated entity descriptions and even web-crawled images to support both recommendation and response generation.
Despite their effectiveness, these context expansion strategies also introduce practical challenges: the resulting inputs are often long, heterogeneous in format, and noisy, which can strain the language understanding capacity of CRS models and lead to diminishing returns. 
Motivated by this limitation, instead of further increasing the amount of external information, we endow CRSs with a screen skim reading capability that enables multi-granularity processing of massive context, improving textual awareness while remaining robust to noisy and lengthy inputs.

\subsection{Visual-centric Token Representation}
Tokenization-based text representation, which segments input text into discrete subword tokens, remains the dominant paradigm in natural language processing.
However, as modern applications increasingly involve long, noisy, and heterogeneous inputs, tokenization can exhibit several limitations, including reduced efficiency on lengthy sequences, sensitivity to typos and irregular formatting, and constraints imposed by a fixed vocabulary~\cite{ali2024tokenizer}.

To mitigate these issues, ``treating text as vision” has emerged as an alternative paradigm~\cite{salesky2021robust,rust2022language,gao2024improving}, where text is rendered as pixels and processed without explicit tokenization.
For example, Rust et al.~\cite{rust2022language} convert text into images and pre-train models by reconstructing masked regions at the pixel level, demonstrating strong robustness and promising cross-lingual transfer.
Motivated by these findings, a growing body of work has developed pre-trained visual-centric representations for text~\cite{chai2024dual,gao2024improving,lotz2023text,xiao2024pixel,xing2025vision}.
For instance, DeepSeek-OCR~\cite{wei2025deepseek} provides an encoder that embeds OCR-rendered text images across multiple resolutions, enabling flexible handling of diverse layouts.

Recognizing that visual-centric and token-based representations offer complementary strengths, recent studies have explored hybrid designs that combine both modalities to better process visually situated text, such as web pages~\cite{lee2023pix2struct}, tables~\cite{xing2025tabledart}, and long documents~\cite{kim2022ocr,hu2025mplug}.
In this work, we take an initial step toward bringing visual-centric representations into CRSs, leveraging them to enhance textual awareness under increasingly complex and lengthy contextual inputs.

\section{Preliminaries}
\noindent\textbf{Notation.} 
Throughout this paper, bold lowercase letters (e.g., $\mathbf{a}$) denote vectors, bold uppercase letters (e.g., $\mathbf{A}$) denote matrices and tensors, and calligraphic letters (e.g., $\mathcal{A}$) denote sets or functions.

\noindent\textbf{Conversational Recommender Systems.}
In a CRS, a conversation $\mathcal{C}$ is a sequence of utterances, i.e., $\mathcal{C}=\{\mathcal{U}_{i}\}_{i=1}^{n_{u}}$, where $n_{u}$ is the number of turns and may vary across conversations. 
Each utterance $\mathcal{U}_{i}$ consists of a sequence of tokens (or words), denoted as $\mathcal{U}_{i}=\{w_{j}\}_{j=1}^{n_{w}}$, where $n_{w}$ is the utterance length.
CRSs typically involve two tasks: response generation and item recommendation. 
For response generation, given the dialogue $\mathcal{C}$, the system generates the next response utterance in natural language.
For recommendation, given the same context $\mathcal{C}$, the system selects a ranked list of recommended items $\mathcal{I}_{rec}$ from the item set $\mathcal{I}$.
Following recent state-of-the-art CRSs~\cite{wei2025mscrs,dao2024broadening}, we incorporate a knowledge graph $\mathcal{G}=\{(e_h, r, e_t)\}$ to model structured relationships among entities mentioned in conversations, where $e_h$, $r$, and $e_t$ denote the head entity, relation, and tail entity, respectively. 
Let $\mathcal{E}$ be the set of all entities in the dataset, and items are a subset of entities, i.e., $\mathcal{I}\subset\mathcal{E}$.

\noindent\textbf{Clarify the Difference from Multi-modal CRSs.}
We emphasize that \modelname is not a multi-modal CRS~\cite{su2024sample} or a visual CRS~\cite{kawamae2025one}. 
Our system operates on textual inputs throughout. 
The ``visual'' signals in \modelname are renderings of text (e.g., screenshot-style representations) introduced solely to obtain a vision-centric encoding of auxiliary textual context, rather than leveraging external visual content.

\begin{figure*}[t]
    \centering
    \includegraphics[width=0.8\textwidth]{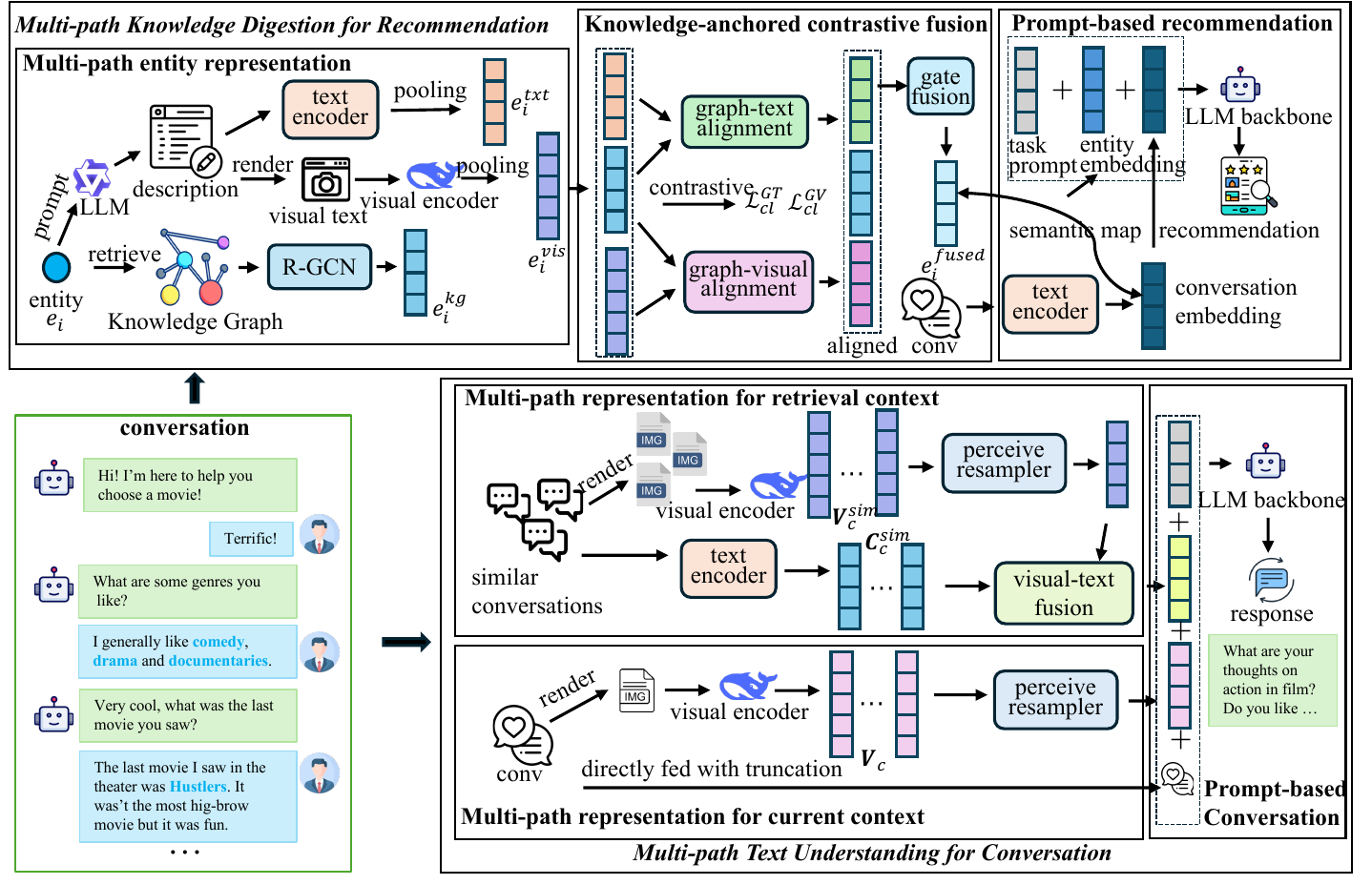}
    \caption{The framework of \modelname.}
    \label{fig_overview}
\end{figure*}

\begin{figure}[t]
    \centering
    \includegraphics[width=0.5\textwidth]{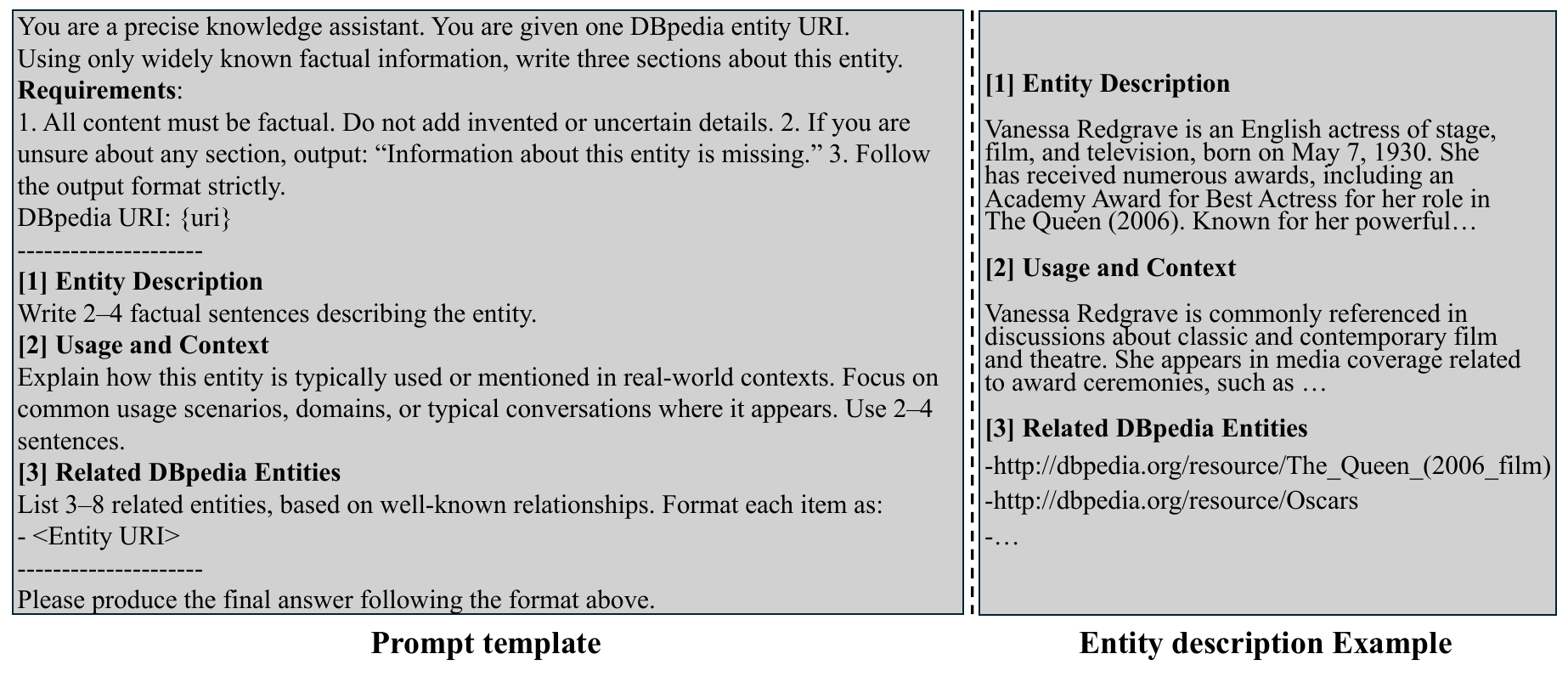}
    \caption{Prompt template and entity description example.}
    \label{fig_template}
\end{figure}

\section{Methodology}
\subsection{Overview}
As discussed above, increasingly rich yet complex textual contexts pose significant challenges for CRSs, as models must contend with longer inputs, heterogeneous formats, and substantial noise.
To address these challenges, we propose \modelname, which incorporates vision-centric text representations to complement conventional text-based representations in CRSs. The core idea is to process contextual information through two complementary pathways. For the most concise and informative portions of the context, we employ a standard text encoder, leveraging its strong semantic modeling and fine-grained reasoning capabilities. In parallel, we utilize a vision-centric encoder to render and embed the entire context as visual tokens. This pathway is naturally more robust to noisy structures and enables coverage of broader contextual information at a coarser semantic granularity. This design is inspired by human reading behavior: when faced with long articles, readers often skim most of the content to gain a high-level understanding, while carefully examining a small set of key sentences to draw reliable conclusions.

Figure~\ref{fig_overview} illustrates the overall architecture of \modelname, which consists of two major modules: Multi-path Knowledge Digestion for Recommendation and Multi-path Text Understanding for Conversation. 
The recommendation module contains three components: 
(1) a multi-path entity representation component that enriches entity information with a knowledge graph and LLM-generated descriptions, and encodes these signals via both textual and visual pathways; 
(2) a knowledge-anchored fusion component that aligns representations from different paths into a shared space and fuses them through contrastive objectives; and 
(3) a prompt learning component that constructs fused prompts to optimize the recommendation task.

The conversation module represents both the retrieved similar conversations and the current dialogue history from textual and visual perspectives. 
It then employs a perceiver sampling mechanism to align and compress heterogeneous token sequences into a unified semantic space. 
Finally, \modelname performs prompt-based response generation conditioned on the fused visual--text prompts.

In the following subsections, we describe the technical details of each component in \modelname.

\subsection{Multi-path Knowledge Digestion for Recommendation}
\subsubsection{Multi-path Entity Representation}
Prior CRS studies have shown that explicitly modeling entities is crucial for accurate recommendations, and thus commonly incorporate a knowledge graph $\mathcal{G}$ to enhance preference reasoning~\cite{zhou2020improving,zhang2023variational}.

\noindent\textbf{Entity Description Generation.}
While $\mathcal{G}$ captures relational structure among entities, it provides limited semantic information about an entity itself.
To enrich entity semantics, \modelname leverages the world knowledge embedded in LLMs.
Specifically, we use $\mathcal{F}_{llm}$ to generate a natural-language description for each entity $e_i$:
\begin{equation}
  \mathcal{T}_{e_i} \leftarrow \mathcal{F}_{llm}(e_i, \mathcal{P}|\mathbf{\Theta}_{llm})
\end{equation}
where $\mathcal{P}$ denotes the prompt template, and $\mathcal{T}_{e_i}$ is the generated description. Figure~\ref{fig_template} shows an example of a prompt and the generated description. Then, given $\mathcal{G}$ and the generated description $\mathcal{T}_{e_i}$, \modelname encodes each entity from three complementary perspectives: (i) knowledge-graph-based encoding, (ii) text-centric encoding, and (iii) vision-centric encoding.

\noindent\textbf{Knowledge-graph-based Entity Encoding.}
Following ~\cite{zhang2023variational,dao2024broadening}, we adopt a relational graph convolutional network (R-GCN)~\cite{schlichtkrull2018modeling} to obtain the structural representation of entity $e_i$ in $\mathcal{G}$:
\begin{equation}
  \mathbf{e}_{i}^{kg(l)} =
  \mathrm{ReLU}\!\left(
  \sum_{r\in\mathcal{R}}
  \sum_{e_j\in\mathcal{N}_r(e_i)}
  \mathbf{W}_{r}^{kg(l)}\mathbf{e}_{j}^{kg(l-1)} + \mathbf{b}_{r}^{kg(l)}
  \right)
\end{equation}
where $\mathbf{e}_{i}^{kg(l)}\in\mathbb{R}^{d_{kg}}$ denotes the $l$-th layer KG representation, $\mathcal{R}$ is the relation set, $\mathcal{N}_r(e_i)$ denotes $e_i$'s neighbors under relation $r$, and $\mathbf{W}_{r}^{kg(l)}$ and $\mathbf{b}_{r}^{kg(l)}$ are trainable parameters.
We aggregate representations across layers to obtain the final KG embedding:
\begin{equation}
  \mathbf{e}_{i}^{kg} = \mathrm{AvgPool}\big(\{\mathbf{e}_{i}^{kg(l)}\}_{l=1}^{L_{kg}}\big)
\end{equation}

\noindent\textbf{Entity Description Encoding via Text Encoder.}
We further encode the generated description $\mathcal{T}_{e_i}$ with a text encoder $\mathcal{F}_{txt}$ to obtain a text-centric embedding $\mathbf{e}_{i}^{txt}\in\mathbb{R}^{d_{txt}}$:
\begin{equation}\label{eq_entity_txt_avg}
  \mathbf{e}_{i}^{txt}=
  \mathrm{AvgPool}\!\left(
  \mathcal{F}_{txt}\big(\mathrm{truncate}(\mathcal{T}_{e_i})|\mathbf{\Theta}_{txt}\big)
  \right)
\end{equation}
Due to the input-length constraint of token-based encoders, we truncate $\mathcal{T}_{e_i}$ to the maximum length supported by $\mathcal{F}_{txt}$.
This pathway provides strong fine-grained semantic understanding, but may omit auxiliary details (e.g., usage scenarios, related entities, and long-tail facts) that appear beyond the truncation boundary.

\noindent\textbf{Entity Description Encoding via Vision-centric Encoder.}
As illustrated in Figure~\ref{fig_template}, entity descriptions can be lengthy and heterogeneous, often containing mixed sections and noisy structures that are difficult for standard token-based encoders to fully digest.
To capture the full description, \modelname introduces a vision-centric pathway.
We first render the original description $\mathcal{T}_{e_i}$ into screenshot-style image(s) using a rendering function:
\begin{equation}
  \mathcal{V}_{e_i} \leftarrow \mathcal{F}_{render}(\mathcal{T}_{e_i})
\end{equation}
where $\mathcal{V}_{e_i}$ may contain multiple images when the description is long.
This rendering preserves character-level patterns, typography, and spatial layout information.
We then apply a pretrained vision encoder $\mathcal{F}_{vis}$ to obtain $L_{e_i}^{vis}$ visual token embeddings:
\begin{equation}
  \{\mathbf{e}_{i,j}^{vis}\}_{j=1}^{L_{e_i}^{vis}} = \mathcal{F}_{vis}(\mathcal{V}_{e_i}|\mathbf{\Theta}_{vis})
\end{equation}
Finally, we summarize the visual tokens via average pooling to produce a single vision-centric entity representation:
\begin{equation}\label{eq_entity_vis_avg}
  \mathbf{e}_{i}^{vis}=\mathrm{AvgPool}\big(\{\mathbf{e}_{i,j}^{vis}\}_{j=1}^{L_{e_i}^{vis}}\big),
  \quad \mathbf{e}_{i}^{vis}\in\mathbb{R}^{d_{vis}}
\end{equation}
This pathway provides coarse-grained yet broader semantics over the entire description, complementing the fine-grained but length-limited text-centric encoding.

\subsubsection{Knowledge-anchored Contrastive Fusion}
After obtaining entity representations from three complementary views, i.e., $\mathbf{e}_{i}^{kg}$, $\mathbf{e}_{i}^{txt}$, and $\mathbf{e}_{i}^{vis}$, we need to fuse them to support recommendations. 
Since these representations are produced by different encoders ($\mathbf{e}_{i}^{kg}$ and the other two embeddings are even generated from different sources), they naturally reside in heterogeneous semantic spaces, making direct fusion suboptimal. 
We therefore first align them in a shared embedding space and perform knowledge-anchored fusion.

\noindent\textbf{Contrastive Alignment of Entity Representations.}
We project the three view-specific embeddings into a common space $\mathbb{R}^{d}$ using MLP-based lightweight projection networks:
\begin{equation}\label{eq_entity_proj}
\begin{aligned}
  \mathbf{e}_{i}^{kg'}  &= \mathcal{F}_{proj}\!\left(\mathbf{e}_{i}^{kg}| \mathbf{\Theta}_{proj}^{kg}\right)\\
  \mathbf{e}_{i}^{txt'} &= \mathcal{F}_{proj}\!\left(\mathbf{e}_{i}^{txt}| \mathbf{\Theta}_{proj}^{txt}\right)\\
  \mathbf{e}_{i}^{vis'} &= \mathcal{F}_{proj}\!\left(\mathbf{e}_{i}^{vis}| \mathbf{\Theta}_{proj}^{vis}\right)
\end{aligned}
\end{equation}
where $\mathbf{\Theta}_{proj}^{*}$ denotes the trainable parameters of each projection layer.
To align semantics across views, we adopt an InfoNCE objective~\cite{oord2018representation} and use the KG representation as an anchor, given its central role in CRS recommendation~\cite{zhou2020improving}. 
Concretely, for a training batch, we treat different views of the same entity as positive pairs, while treating other entities in the batch as negatives:
\begin{equation}\label{eq_entity_cl}
\begin{aligned}
  \mathcal{L}_{kg\text{-}txt}^{cl} &=
  - \sum_{e_{i}\in \mathcal{I}_{b}}
  \log
  \frac{
    \exp\!\left(\mathbf{e}_{i}^{kg'\top}\mathbf{e}_{i}^{txt'}/\tau\right)
  }{
    \sum_{e_{j}\in \mathcal{I}_{b}}
    \exp\!\left(\mathbf{e}_{i}^{kg'\top}\mathbf{e}_{j}^{txt'}/\tau\right)
  }\\
  \mathcal{L}_{kg\text{-}vis}^{cl} &=
  - \sum_{e_{i}\in \mathcal{I}_{b}}
  \log
  \frac{
    \exp\!\left(\mathbf{e}_{i}^{kg'\top}\mathbf{e}_{i}^{vis'}/\tau\right)
  }{
    \sum_{e_{j}\in \mathcal{I}_{b}}
    \exp\!\left(\mathbf{e}_{i}^{kg'\top}\mathbf{e}_{j}^{vis'}/\tau\right)
  }
\end{aligned}
\end{equation}
where $\mathcal{I}_{b}$ denotes the set of entities appearing in the conversations of the current batch, and $\tau$ is a temperature hyper-parameter.
This contrastive alignment encourages representations of the same entity across different views to be close in the shared space, while pushing apart representations of different entities.

\noindent\textbf{Knowledge-anchored Fusion.}
Let $\mathcal{E}_{c}$ be the set of entities mentioned in conversation $\mathcal{C}$, and let the corresponding aligned entity sequences be
$\mathbf{E}_{c}^{kg}$, $\mathbf{E}_{c}^{txt}$, and $\mathbf{E}_{c}^{vis}$ (each constructed by stacking the projected embeddings from E.q.~\ref{eq_entity_proj}). 
To capture fine-grained interactions between the KG view and the two auxiliary views, \modelname applies bidirectional multi-head cross-attention, using the KG embedding as the anchor:
\begin{equation}\label{eq_entity_cross_attn}
\begin{aligned}
  \mathbf{E}_{c}^{txt'} &= \mathrm{CrossAttn}(\mathbf{E}_{c}^{kg}, \mathbf{E}_{c}^{txt})\\
  \mathbf{E}_{c}^{vis'} &= \mathrm{CrossAttn}(\mathbf{E}_{c}^{kg}, \mathbf{E}_{c}^{vis})
\end{aligned}
\end{equation}
where $\mathrm{CrossAttn}(\mathbf{Q},\mathbf{K})$ denotes multi-head cross-attention with $\mathbf{Q}$ as queries and $\mathbf{K}$ as keys and values.

Finally, we fuse the KG features with the cross-attended textual and visual features via an adaptive gating mechanism, which allows the model to emphasize different views depending on the conversation, as different conversation has different focuses:
\begin{equation}
\begin{aligned}
  \mathbf{g}_{txt} &= \sigma\!\left(\mathbf{E}_{c}^{txt'}\mathbf{W}_{txt}^{gate}+\mathbf{b}_{txt}^{gate}\right)\\
  \mathbf{g}_{vis} &= \sigma\!\left(\mathbf{E}_{c}^{vis'}\mathbf{W}_{vis}^{gate}+\mathbf{b}_{vis}^{gate}\right)\\
  \mathbf{E}_{c}^{fuse} &=
  \frac{
    \mathbf{E}_{c}^{kg}
    + \mathbf{g}_{txt}\odot \mathbf{E}_{c}^{txt'}
    + \mathbf{g}_{vis}\odot \mathbf{E}_{c}^{vis'}
  }{
    \mathbf{1} + \mathbf{g}_{txt} + \mathbf{g}_{vis}
  }
\end{aligned}
\end{equation}
where $\sigma(\cdot)$ is the sigmoid function, $\odot$ denotes element-wise multiplication, and $\mathbf{W}_{*}^{gate}, \mathbf{b}_{*}^{gate}$ are trainable parameters.
This formulation yields a normalized, conversation-adaptive fusion that integrates structured KG signals with fine-grained textual semantics and coarse-grained yet broader vision-centric cues.

\subsubsection{Prompt Training for Recommendation}
Following recent practice in LLM-based CRSs~\cite{wei2025mscrs,dao2024broadening,wang2022towards}, we adopt prompt learning~\cite{radford2018improving} to efficiently adapt an LLM for the recommendation task. 

\noindent\textbf{Recommendation Prompt.}
We construct the recommendation prompt $\mathbf{P}_{rec}$ by concatenating (i) fused entity representations, (ii) a trainable soft prompt, and (iii) the dialogue context:
\begin{equation}
  \mathbf{P}_{rec} = [\mathbf{E}^{final}_{c}, \mathbf{W}^{rec}, \mathcal{C}]
\end{equation}
where $\mathbf{W}^{rec}$ denotes trainable soft prompt embeddings for recommendation.
$\mathbf{E}^{final}_{c}$ is obtained by injecting the fused entity sequence $\mathbf{E}^{fuse}_{c}$ into the conversation representation via cross-attention, so that entity semantics can interact with the dialogue context.

After feeding $\mathbf{P}_{rec}$ into the backbone LLM, we use the hidden state of the last token, denoted as $\hat{\mathbf{t}}_{c}$, as the user preference representation for item retrieval, computing the matching score between $\hat{\mathbf{t}}_{c}$ and each candidate item embedding $\mathbf{e}^{fuse}_{i}$.

\noindent\textbf{Stage I: Preference--Entity Alignment Pretraining.}
Since $\hat{\mathbf{t}}_{c}$ is produced in the backbone LLM's representation space while $\mathbf{e}^{fuse}_{i}$ originates from the fused KG-text-vision pathways, there can be a non-negligible semantic gap between them. 
We therefore introduce a pretraining stage that aligns these spaces by training $\hat{\mathbf{t}}_{c}$ to retrieve the entities mentioned in the current conversation, inspired by~\cite{dao2024broadening,deng2021unified}.
Formally, for a batch of conversations $\mathcal{C}_{b}$, we compute:
\begin{equation}
\begin{aligned}
  \hat{r}_{ci} &= \mathrm{Softmax}\!\left(\hat{\mathbf{t}}_{c}^{\top}\mathbf{e}^{fuse}_{i}\right)\\
  \mathcal{L}^{pre} &=
  -\sum_{\mathcal{C}\in \mathcal{C}_{b}}
  \sum_{i\in \mathcal{E}_{c}}
  \log(\hat{r}_{ci})
  \;+\; \alpha \mathcal{L}_{kg\text{-}txt}^{cl}
  \;+\; \beta \mathcal{L}_{kg\text{-}vis}^{cl}
\end{aligned}
\end{equation}
where $\mathcal{E}_{c}$ is the set of entities appearing in conversation $\mathcal{C}$, and $\alpha,\beta$ control the strength of the contrastive alignment losses defined in E.q.~\ref{eq_entity_cl}.

\noindent\textbf{Stage II: Recommendation Fine-tuning.}
Starting from the pretrained parameters, we further fine-tune \modelname for item recommendation by predicting the ground-truth recommended items. 
The recommendation objective is defined as:
\begin{equation}
\begin{aligned}
  \mathcal{L}^{rec} &=
  -\sum_{\mathcal{C}\in \mathcal{C}_{b}}
  \sum_{i\in \mathcal{I}}
  r_{ci}\log(\hat{r}_{ci})
  \;+\; \alpha \mathcal{L}_{kg\text{-}txt}^{cl}
  \;+\; \beta \mathcal{L}_{kg\text{-}vis}^{cl}
\end{aligned}
\end{equation}
where $r_{ci}$ is the ground-truth label indicating whether item $i$ is a target recommendation in conversation $\mathcal{C}$, and $\mathcal{I}$ denotes the full item set.

\subsection{Multi-path Text Understanding for Conversation}
Following recent LLM-based CRSs~\cite{wei2025mscrs,dao2024broadening}, \modelname leverages both the current dialogue context and a set of retrieved similar conversations as augmented evidence for response generation. Since conversations are often long yet contain sparse preference signals, \modelname models them through complementary text-centric and vision-centric representations to improve robustness and coverage. Importantly, retrieved contexts and the current dialogue play different roles: the current conversation directly drives response generation and is thus more critical, whereas retrieved conversations serve as auxiliary evidence. Accordingly, we adopt different processing strategies. Retrieved conversations are transformed into soft prompts before being injected into the backbone LLM, while the current conversation is fed to the backbone in its tokenized form with length limitation truncation and complemented with an additional vision-centric soft prompt to provide a broader, skim-reading view of the full dialogue history.

\subsubsection{Multi-path Representation for Retrieved Conversation}
Following~\cite{wei2025mscrs}, we retrieve $K^{sim}$ most similar conversations, denoted as $\mathcal{C}_{c}^{sim}=\{\mathcal{C}_{j}\}_{j=1}^{K^{sim}}$.
For the retrieved context, we construct both text-centric and vision-centric token representations:
\begin{equation}
\begin{aligned}
  \mathbf{C}_{c}^{sim} &= \mathcal{F}_{txt}\!\left(\mathrm{truncate}(\mathcal{C}_{c}^{sim}) | \mathbf{\Theta}_{txt}\right)\\
  \mathbf{V}_{c}^{sim} &= \mathcal{F}_{vis}\!\left(\mathcal{V}_{c}^{sim} | \mathbf{\Theta}_{vis}\right)
\end{aligned}
\end{equation}
where $\mathcal{V}_{c}^{sim}$ denotes screenshot-style images rendered from $\mathcal{C}_{c}^{sim}$.
Because $\mathbf{C}_{c}^{sim}$ and $\mathbf{V}_{c}^{sim}$ come from different encoders, they may differ in both token length and embedding dimension.
Unlike in entity representations that we simply use average pooling to overcome the token number difference(E.q.~\ref{eq_entity_txt_avg} and~\ref{eq_entity_vis_avg}) and use MLP for dimension alignment (E.q.~\ref{eq_entity_proj}), we employ a trainable Perceiver Resampler~\cite{alayrac2022flamingo} (with projection layers) to compress each sequence into a fixed-length matrix in a shared space.
This is because conversations are vital for response generation, while the entity description is auxiliary information for recommendation.
The formal presentation is as follows:
\begin{equation}
\begin{aligned}
  \overline{\mathbf{C}}_{c}^{sim} &= \mathcal{F}_{perceiver}\!\left(\mathbf{C}_{c}^{sim} | \mathbf{\Theta}_{pr}^{txt}\right)\\
  \overline{\mathbf{V}}_{c}^{sim} &= \mathcal{F}_{perceiver}\!\left(\mathbf{V}_{c}^{sim} | \mathbf{\Theta}_{pr}^{vis}\right)
\end{aligned}
\end{equation}
where $\overline{\mathbf{C}}_{c}^{sim}, \overline{\mathbf{V}}_{c}^{sim}\in\mathbb{R}^{L^{r}\times d}$, and $\mathbf{\Theta}_{pr}^{txt}, \mathbf{\Theta}_{pr}^{vis}$ are trainable parameters for perceiver resampler.
To enable cross-view interaction, we further apply cross-attention to fuse the two pathways and obtain the retrieved-context prompt:
\begin{equation}
  \widetilde{\mathbf{C}}_{c}^{sim} = \mathrm{CrossAttn}\!\left(\overline{\mathbf{C}}_{c}^{sim}, \overline{\mathbf{V}}_{c}^{sim}\right)
\end{equation}

\subsubsection{Multi-path Representation for the Current Context}
For the current conversation, the original tokenized dialogue $\mathcal{C}$ will be fed into the LLM backbone directly. 
Besides, we construct an additional vision-centric soft prompt that captures the global context in a broad skim-reading manner.
Specifically, we render the current conversation into an image set $\mathcal{V}_{c}$ and encode it with $\mathcal{F}_{vis}$, followed by the Perceiver Resampler to match the pre-defined prompt length and backbone LLM's hidden size:
\begin{equation}
\begin{aligned}
  \mathbf{V}_{c} &= \mathcal{F}_{vis}\!\left(\mathcal{V}_{c}|\mathbf{\Theta}_{vis}\right)\\
  \overline{\mathbf{V}}_{c} &= \mathcal{F}_{perceiver}\!\left(\mathbf{V}_{c}|\mathbf{\Theta}_{pc}^{vis}\right)
\end{aligned}
\end{equation}
where $\overline{\mathbf{V}}_{c}\in\mathbb{R}^{\gamma\times d}$, $\gamma$ is the pre-defined visual soft-prompt length for the current conversation, and $\mathbf{\Theta}_{pc}^{vis}$ are trainable parameters.

\subsubsection{Prompt Training for Conversation}
We construct the final prompt for response generation as:
\begin{equation}
  \mathbf{P}_{conv} = [\widetilde{\mathbf{C}}_{c}^{sim},\; \overline{\mathbf{V}}_{c},\; \mathbf{W}^{conv},\; \mathcal{C}],
\end{equation}
where $\mathbf{W}^{conv}$ denotes trainable prompt embeddings for the conversation task.
Given a training batch $\mathcal{C}_{b}$, we optimize the model using the standard autoregressive negative log-likelihood:
\begin{equation}
  \mathcal{L}^{conv} =
  - \sum_{\mathcal{C}\in \mathcal{C}_{b}}
  \sum_{j=1}^{l_{c}}
  \log
  \mathcal{P}_{gen}\!\left(w_{c,j}\mid w_{c,<j}, \mathbf{P}_{conv}\right),
\end{equation}
where $l_{c}$ is the length of the ground-truth response, and $\mathcal{P}_{gen}$ denotes the generation probability conditioned on the prompt and the preceding tokens.

\section{Experiments}
\subsection{Experimental Setup}
\subsubsection{Datasets}
Following~\cite{wei2025mscrs,dao2024broadening,zou2022improving}, we conduct extensive experiments on two widely used conversational recommendation datasets: ReDial~\cite{li2018towards} and INSPIRED~\cite{hayati2020inspired}.
These two datasets are both for movie recommendation, but ReDial is curated through Amazon Mechanical Turk, while INSPIRED focuses on persuasive outcomes with social strategies.
The statistics are in Table~\ref{tb_ds_statistics}.

\begin{table}[!htbp]
  \centering
  \caption{Statistics of datasets.}\label{tb_ds_statistics}
  \resizebox{\linewidth}{!}{
\begin{tabular}{l|cccc}
\toprule
\textbf{Dataset}  & \textbf{\# users} & \textbf{\# conversation} & \textbf{\# utterances} & \textbf{\# entities/items} \\ \hline
\textbf{ReDial}   & 956               & 10,006                   & 182,150                & 64,364/6924                \\
\textbf{INSPIRED} & 1,482             & 1,001                    & 35,811                 & 17,321/1,123               \\ \bottomrule
\end{tabular}}
\end{table}

\begin{table*}[!htbp]
\caption{Comparison of recommendation performance on ReDial and INSPIRED. The bold values denote the best performances.}
\label{tb_rec}
\centering
\resizebox{\textwidth}{!}{
\begin{tabular}{c|ccc|cc|cc|ccc|cc|cc}
\toprule
 & \multicolumn{7}{c|}{\textbf{ReDial}} & \multicolumn{7}{c}{\textbf{INSPIRED}} \\
\cmidrule(lr){2-8} \cmidrule(lr){9-15}
\multirow{1}{*}{\textbf{Model}} & \multicolumn{3}{c|}{\textbf{Recall}} & \multicolumn{2}{c|}{\textbf{NDCG}} & \multicolumn{2}{c|}{\textbf{MRR}} & \multicolumn{3}{c|}{\textbf{Recall}} & \multicolumn{2}{c|}{\textbf{NDCG}} & \multicolumn{2}{c}{\textbf{MRR}} \\
\cmidrule(lr){2-4} \cmidrule(lr){5-6} \cmidrule(lr){7-8} \cmidrule(lr){9-11} \cmidrule(lr){12-13} \cmidrule(lr){14-15}
 & \textbf{@1} & \textbf{@10} & \textbf{@50} & \textbf{@10} & \textbf{@50} & \textbf{@10} & \textbf{@50} & \textbf{@1} & \textbf{@10} & \textbf{@50} & \textbf{@10} & \textbf{@50} & \textbf{@10} & \textbf{@50} \\
\midrule
Popularity     & 0.011 & 0.053 & 0.183 & 0.029 & 0.057 & 0.021 & 0.027 & 0.031 & 0.155 & 0.322 & 0.085 & 0.122 & 0.064 & 0.071 \\
TextCNN   & 0.010 & 0.066 & 0.187 & 0.033 & 0.059 & 0.023 & 0.028 & 0.025 & 0.119 & 0.245 & 0.066 & 0.094 & 0.050 & 0.056 \\
BERT       & 0.027 & 0.142 & 0.307 & 0.075 & 0.112 & 0.055 & 0.063 & 0.049 & 0.189 & 0.322 & 0.112 & 0.141 & 0.088 & 0.095 \\
\midrule
ReDial    & 0.010 & 0.065 & 0.182 & 0.034 & 0.059 & 0.024 & 0.029 & 0.009 & 0.048 & 0.213 & 0.023 & 0.059 & 0.015 & 0.023 \\
KBRD     & 0.033 & 0.150 & 0.311 & 0.083 & 0.118 & 0.062 & 0.070 & 0.042 & 0.135 & 0.236 & 0.088 & 0.109 & 0.073 & 0.077 \\
KGSF      & 0.035 & 0.175 & 0.367 & 0.094 & 0.137 & 0.070 & 0.079 & 0.051 & 0.132 & 0.239 & 0.092 & 0.114 & 0.079 & 0.083 \\
TREA    & 0.045 & 0.204 & 0.403 & 0.114 & 0.158 & 0.087 & 0.096 & 0.047 & 0.146 & 0.347 & 0.095 & 0.132 & 0.076 & 0.087 \\
VRICR    & 0.054 & 0.244 & 0.406 & 0.138 & 0.174 & 0.106 & 0.114 & 0.043 & 0.141 & 0.336 & 0.091 & 0.134 & 0.075 & 0.085 \\
UNICRS   & 0.065 & 0.241 & 0.423 & 0.143 & 0.183 & 0.113 & 0.125 & 0.085 & 0.230 & 0.398 & 0.149 & 0.187 & 0.125 & 0.133 \\
DCRS & 0.076& 0.253 & 0.439 & 0.154 & 0.195 & 0.123 & 0.132 & 0.093 & 0.226 & 0.414 & 0.153 & 0.192 & 0.130 & 0.137 \\
MSCRS  & 0.081 & 0.264 & 0.451 & 0.161 & 0.201 & 0.128 & 0.136 & 0.096 & 0.257 & 0.425 & 0.168 & 0.202 & 0.140 & 0.148 \\\midrule
\textbf{\modelname}  & \textbf{0.083} & \textbf{0.271} & \textbf{0.453} & \textbf{0.163} & \textbf{0.207} & \textbf{0.132} & \textbf{0.139} & \textbf{0.098}& \textbf{0.301} & \textbf{0.457} & \textbf{0.193} & \textbf{0.228} & \textbf{0.159} & \textbf{0.167}\\
Improve.(\%) & 2.46 & 2.65 & 0.44 & 1.24 & 2.98 & 3.12 & 2.20 & 2.08 & \textbf{17.12} & 7.52 & 14.88 & 12.87 & 13.57 & 12.83\\
\bottomrule
\end{tabular}
}
\end{table*}

\subsubsection{Evaluation Metrics}
In the experiments, we will evaluate both recommendation and conversation performance. All the evaluation protocols follow the previous works~\cite{wei2025mscrs,dao2024broadening,deng2021unified,wang2022towards}.
For recommendation task, we utilize Recall@K ($K\in \{1, 10, 50\}$), NDCG@K ($K\in \{10, 50\}$), and MRR@K ($K \in \{10, 50\}$). Note that we do not report NDCG@1 or MRR@1, as both are equivalent to Recall@1.

The conversation quality is assessed with both automatic and human evaluations.
The automatic metrics include BELU-N ($N\in \{2,3\}$), ROUGE-N ($N\in \{2, L\}$), and Distinct-N ($N\in \{2,3,4\}$).
For human evaluation, following~\cite{wei2025mscrs,dao2024broadening}, we randomly sampled $50$ responses from each model and asked two annotators to score them on Fluency and Informativeness with $0\leq score \leq 1$, and averaged all the scores.

\subsubsection{Baselines}
Following~\cite{wei2025mscrs,dao2024broadening,an2025beyond}, we select the following competitive baselines for performance comparison:

    \noindent\textbf{Traditional recommenders and General pretrained LLMs: } 
    (1) \textbf{Popularity}: ranks candidate items by their historical occurrence frequency in the training corpus; (2) \textbf{TextCNN}~\cite{kim2014convolutional}: applies convolutional neural networks to encode the dialogue context for item ranking/prediction.
        (3) \textbf{BERT}~\cite{devlin2019bert}: a pre-trained bidirectional Transformer; we fine-tune it to predict a set of candidate items given the conversation history.
        (4) \textbf{GPT-2}~\cite{radford2019language}: an autoregressive language model serving as a strong text generation baseline.
        (5) \textbf{DialoGPT}~\cite{zhang2020dialogpt}: a dialogue-oriented GPT model pre-trained on large-scale conversational data for response generation.
        (6) \textbf{BART}~\cite{lewis2020bart}: a denoising sequence-to-sequence pre-trained model for natural language generation. 

    \noindent\textbf{State-of-the-art CRS methods:}
    (1) \textbf{ReDial}~\cite{li2018towards}: an early CRS that combines an auto-encoder-based recommender with a hierarchical RNN dialogue generator.
        (2) \textbf{KBRD}~\cite{chen2019towards}: a knowledge-enhanced CRS that leverages entity graph propagation to improve recommendation and response generation.
        (3) \textbf{KGSF}~\cite{zhou2020improving}: integrates both entity-oriented and word-oriented graphs to enrich semantic representations for CRS.
        (4) \textbf{TREA}~\cite{li2023trea}: introduces structured (tree-style) reasoning to jointly support recommendation and dialogue generation.
        (5) \textbf{UniCRS}~\cite{wang2022towards}: unifies recommendation and generation within a prompt-learning paradigm.
        (6) \textbf{VRICR}~\cite{zhang2023variational}: improves reasoning with variational Bayesian inference to better handle incomplete knowledge for CRS.
        (7) \textbf{DCRS}~\cite{dao2024broadening}: augments the dialogue context with retrieved demonstration conversations to strengthen understanding and generation.
        (8) \textbf{MSCRS}~\cite{wei2025mscrs}: further incorporates LLM-generated texts and web-crawled images to augment entity information for recommendation and conversation. It is the current state-of-the-art CRS.

\subsubsection{Implementation Details}  %
The backbone LLM is DialoGPT-small~\cite{zhang2020dialogpt}, 
which follows recent CRS studies~\cite{wei2025mscrs,dao2024broadening,deng2021unified} to ensure that the observed improvements primarily stem from our methodological design rather than upgrading backbone models.
We instantiate the textual encoder with RoBERTa-base~\cite{liu2019roberta}.
The $\mathcal{F}_{llm}$ is initialized with Qwen3-8B~\cite{yang2025qwen3}, while for the vision-centric pathway, we adopt the DeepSeek-OCR encoder~\cite{wei2025deepseek}, considering their availability and performance.
All these pretrained models are frozen during \modelname's training.
For the knowledge graph, we use a 1-layer R-GCN to process it. 
Unless otherwise specified, all cross-attention modules in \modelname use a single layer, and the perceiver resampler is also configured with one layer. 
We tune the soft prompt length in the range of $[10,70]$ for both the recommendation and conversation tasks. 
The contrastive coefficients $\alpha$ and $\beta$ are selected from $1 \times 10^{-4}$ to $1\times10^{-2}$.
We optimize all models with Adam~\cite{kingma2014adam}, with the learning rate searched from $1 \times 10^{-4}$ to $1 \times 10^{-3}$.

For baselines, we keep consistent with their original results reported in their papers unless the reproduced results deviate substantially.
For \modelname, we run five independent trials with different random seeds and report the average results. 
All improvements are statistically significant with $p<0.05$.

\begin{table*}[tbp]
\centering
\setlength{\tabcolsep}{4pt} %
\caption{Comparison of conversation performance on Redial and INSPIRED datasets with automatic metrics. The bold value indicates the best performance.}
\label{tb_cov}
\begin{tabular}{c|cc|cc|ccc|cc|cc|ccc}
\toprule
 & \multicolumn{7}{c|}{\textbf{ReDial}} & \multicolumn{7}{c}{\textbf{INSPIRED}} \\
\cmidrule(lr){2-8} \cmidrule(lr){9-15}
\multirow{1}{*}{\textbf{Model}} & \multicolumn{2}{c|}{\textbf{BLEU}} & \multicolumn{2}{c|}{\textbf{ROUGE}} & \multicolumn{3}{c|}{\textbf{DIST}} & \multicolumn{2}{c|}{\textbf{BLEU}} & \multicolumn{2}{c|}{\textbf{ROUGE}} & \multicolumn{3}{c}{\textbf{DIST}} \\

\cmidrule(lr){2-3} \cmidrule(lr){4-5} \cmidrule(lr){6-8} \cmidrule(lr){9-10} \cmidrule(lr){11-12} \cmidrule(lr){13-15}
 & \textbf{-2} & \textbf{-3} & \textbf{-2} & \textbf{-L} & \textbf{-2} & \textbf{-3} & \textbf{-4} &\textbf{-2} & \textbf{-3} & \textbf{-2} & \textbf{-L} & \textbf{-2} & \textbf{-3}& \textbf{-4}  \\
\midrule
    DialogGPT & 0.041 & 0.021 & 0.054 & 0.258 & 0.436 & 0.632 & 0.771 & 0.031 & 0.014 & 0.041 & 0.207 & 1.954 & 2.750 & 3.235 \\
    GPT-2& 0.031 & 0.013 & 0.041 & 0.244 & 0.405 & 0.603 & 0.757 & 0.026 & 0.011 & 0.034 & 0.212 & 2.119 & 3.084 & 3.643 \\
    BART & 0.024 & 0.011 & 0.031 & 0.229 & 0.432 & 0.615 & 0.705 & 0.018 & 0.008 & 0.025 & 0.208 & 1.920 & 2.501 & 2.670 \\
    \midrule
    ReDial & 0.004 & 0.001 & 0.021 & 0.187 & 0.058 & 0.204 & 0.442 & 0.001 & 0.000 & 0.004 & 0.168 & 0.359 & 1.043 & 1.760 \\
    KBRD & 0.038 & 0.018 & 0.047 & 0.237 & 0.070 & 0.288 & 0.488 & 0.021 & 0.007 & 0.029 & 0.218 & 0.416 & 1.375 & 2.320 \\
    KGSF& 0.030 & 0.012 & 0.039 & 0.244 & 0.061 & 0.278 & 0.515 & 0.023 & 0.007 & 0.031 & 0.228 & 0.418 & 1.496 & 2.790\\
    VRICR& 0.021 & 0.008 & 0.027 & 0.137 & 0.107 & 0.286 & 0.471 & 0.011 & 0.001 & 0.025 & 0.187 & 0.853 & 1.801 & 2.827\\
    TREA& 0.022 & 0.008 & 0.039 & 0.175 & 0.242 & 0.615 & 1.176 & 0.013 & 0.002 & 0.027 & 0.195 & 0.958 & 2.565 & 3.411\\
    UNICRS& 0.045 & 0.021 & 0.058 & 0.285 & 0.433 & 0.748 & 1.003 & 0.022 & 0.009 & 0.029 & 0.212 & 2.686 & 4.343 & 5.520 \\
    DCRS& 0.048 & 0.024 & 0.063 & 0.285 & 0.779 & 1.173 & 1.386 & 0.033 & 0.014 & 0.045 & 0.229 & 3.950 & 5.729 & 6.233  \\
   MSCRS & 0.043 & 0.017 & 0.058 & 0.287 & 0.770 & 1.112 & 1.310 &0.035 & 0.016  & 0.047 & 0.234 & 2.282 & 3.331 & 3.985\\    \midrule
   \textbf{\modelname} & \textbf{0.051} & \textbf{0.025} & \textbf{0.066} & \textbf{0.295} & \textbf{0.781} & \textbf{1.192} & \textbf{1.419} &\textbf{0.041} & \textbf{0.019}  & \textbf{0.054} & \textbf{0.244} & \textbf{3.997} & \textbf{5.824} & \textbf{6.401}\\
   Improve.(\%) & 6.25 & 4.16 & 4.76 & 2.78 & 0.26 & 1.62 & 2.38 & 17.14 & \textbf{18.75}  & 14.89 & 4.27 & 1.19 & 1.66 & 2.69\\
    \bottomrule
\end{tabular}
\end{table*}

\subsection{Performance on Recommendation Task}
Table~\ref{tb_rec} reports the recommendation performance of \modelname and representative CRS baselines on the ReDial and INSPIRED datasets. 
Overall, \modelname achieves state-of-the-art results across all evaluated metrics.

In particular, compared with the strongest prior baseline MSCRS, \modelname yields substantial improvements on INSPIRED, achieving $17.12\%$ higher Recall@10, $14.88\%$ higher NDCG@10, and $13.57\%$ higher MRR@10. 
On ReDial, \modelname also consistently outperforms MSCRS, with around a $2\%$ gain in both Recall@10 and MRR@10. 
Notably, \modelname relies solely on textual information, whereas MSCRS additionally requires crawling related images from the Internet, which may hinder deployment in scenarios where external image retrieval is unavailable or undesirable.

We further observe that the improvement margins on INSPIRED are larger than those on ReDial. 
This difference is likely driven by dataset characteristics: ReDial focuses on movie recommendation and contains denser and more explicit preference cues, whereas INSPIRED is more persuasion- and explanation-oriented, with user preference signals that are comparatively sparse. 
As a result, enriching entity information and improving the digestion of auxiliary context via multi-path processing brings larger benefits on INSPIRED than on ReDial, as the latter already has rich collaborative information.

\begin{table}[!htbp]
\centering
\caption{Human evaluation for the conversation task.}
\label{tb_human_language}
\begin{tabular}{lcc}
    \toprule
    \textbf{Models} & \textbf{Fluency} & \textbf{Informativeness} \\
    \midrule
    DialoGPT & 0.67 & 0.59 \\
    KBRD & 0.61 & 0.52 \\
    UNICRS & 0.74 & 0.63\\
    DCRS & 0.79 & 0.67 \\
    MSCRS & 0.70 & 0.63 \\
    \midrule
    \textbf{\modelname} & \textbf{0.82} & \textbf{0.73} \\
    \bottomrule
\end{tabular}
\end{table}

\subsection{Performance on Conversation Task}
To assess response quality, we conduct both automatic and human evaluations. 
Table~\ref{tb_cov} summarizes the results on automatic metrics. 
BLEU and ROUGE measure the overlap between generated responses and the reference responses at the n-gram level, reflecting surface-level consistency. 
DIST captures lexical diversity, which is complementary to BLEU/ROUGE but does not necessarily indicate factual correctness or relevance.

Overall, \modelname consistently outperforms all baselines, including recent state-of-the-art CRS models. 
Compared with the strongest baseline on ReDial, \modelname improves BLEU-2 by $6.25\%$. 
On INSPIRED, \modelname achieves an $18.75\%$ gain on BLEU-3, representing a substantial improvement in generation quality. 
We attribute these gains to our multi-path text understanding design, which strengthens the model's ability to process complex and expanded contexts. 
Moreover, relative to the backbone generator DialoGPT, \modelname yields large improvements on ROUGE-2, with gains of $22\%$ on ReDial and $31\%$ on INSPIRED, further validating the effectiveness of the overall system design.

To further validate response quality beyond automatic metrics, we perform human evaluation following prior CRS protocols~\cite{wei2025mscrs,dao2024broadening} on ReDial. 
We compare \modelname with the top-5 baselines that achieve the strongest conversation performance according to automatic metrics. 
As shown in Table~\ref{tb_human_language}, \modelname consistently receives the highest scores in both fluency and informativeness.

\begin{table}[!htbp]
  \centering
  \caption{Ablation studies of two text understanding paths on the recommendation task.}\label{tb_ablation_rec}
  \resizebox{\linewidth}{!}{
\begin{tabular}{l|ccc|ccc}
\toprule
             & \multicolumn{3}{c|}{\textbf{ReDial}}            & \multicolumn{3}{c}{\textbf{INSPIRED}}          \\ \hline
   Metrics@10          & \textbf{Recall} & \textbf{NDCG} & \textbf{MRR} & \textbf{Recall} & \textbf{NDCG} & \textbf{MRR} \\ \hline
\modelname         & 0.271           & 0.163         & 0.132        & 0.301           & 0.193         & 0.159        \\
-entity text path   & 0.255           & 0.154         & 0.124        & 0.262           & 0.163         & 0.136        \\
-entity visual path & 0.260           & 0.158         & 0.127        & 0.281           & 0.176         & 0.143        \\ \bottomrule
\end{tabular}}
\end{table}

\begin{table}[!htbp]
  \centering
  \caption{Ablation studies of two text understanding paths on the conversation task.}\label{tb_ablation_conv}
\begin{tabular}{l|cccc}
\toprule
                       & \multicolumn{2}{c}{\textbf{ReDial}} & \multicolumn{2}{c}{\textbf{INSPIRED}} \\ \hline
Metrics-2              & \textbf{BLEU}    & \textbf{ROUGE}   & \textbf{BLEU}     & \textbf{ROUGE}    \\ \hline
\modelname                   & 0.051            & 0.066            & 0.041             & 0.054             \\
-retrieval text path   & 0.048            & 0.064            & 0.039             & 0.051            \\
-retrieval visual path & 0.049            & 0.065            & 0.039             & 0.052            \\
-current  visual path & 0.046            & 0.063            & 0.038             & 0.050            \\ \bottomrule  
\end{tabular}
\end{table}

\subsection{Ablation Study}
We conduct ablation studies to examine the contributions of the vision-centric and textual encoding pathways in \modelname for both the recommendation and conversation tasks.

\subsubsection{Ablation of multi-path understanding for recommendation}
For the recommendation task, the multi-path design is primarily applied to entity description. 
In Table~\ref{tb_ablation_rec}, we ablate the textual pathway (\textit{-entity text path}) and the visual pathway (\textit{-entity visual path}) for entity encoding, respectively. 
The results show that removing either pathway consistently degrades performance, indicating that both paths contribute complementary signals to preference modeling. 
We further observe that retaining only the visual pathway performs worse than retaining only the textual pathway. 
A plausible explanation is that, while vision-centric tokens are good at dealing with long contexts, they may still lose fine-grained semantic details compared with standard token-based text encoders, which are better suited for capturing precise entity semantics.

\subsubsection{Ablation of multi-path understanding for conversation}
For the conversation task, \modelname applies multi-path understanding to two sources of context: retrieved similar conversations and the current dialogue history. 
Since our response generator is an LLM, we keep the textual pathway for the current conversation as the default backbone, as removing it would fundamentally break the generation process and lead to uninformative comparisons. 
We therefore ablate (i) the visual pathway for retrieved conversations (\textit{-retrieved visual path}), (ii) the textual pathway for retrieved conversations (\textit{-retrieved text path}), and (iii) the visual pathway for the current conversation (\textit{-current visual path}). 

As reported in Table~\ref{tb_ablation_conv}, removing any of these pathways results in noticeable performance degradation, validating the effectiveness of multi-granularity text understanding. 
Among the ablations, dropping the current visual pathway causes the largest decline, likely because it directly weakens \modelname's ability to capture salient cues from the ongoing dialogue under noisy or lengthy contexts, which subsequently affects response generation.

\begin{figure}[t]
    \centering
    \includegraphics[width=0.47\textwidth]{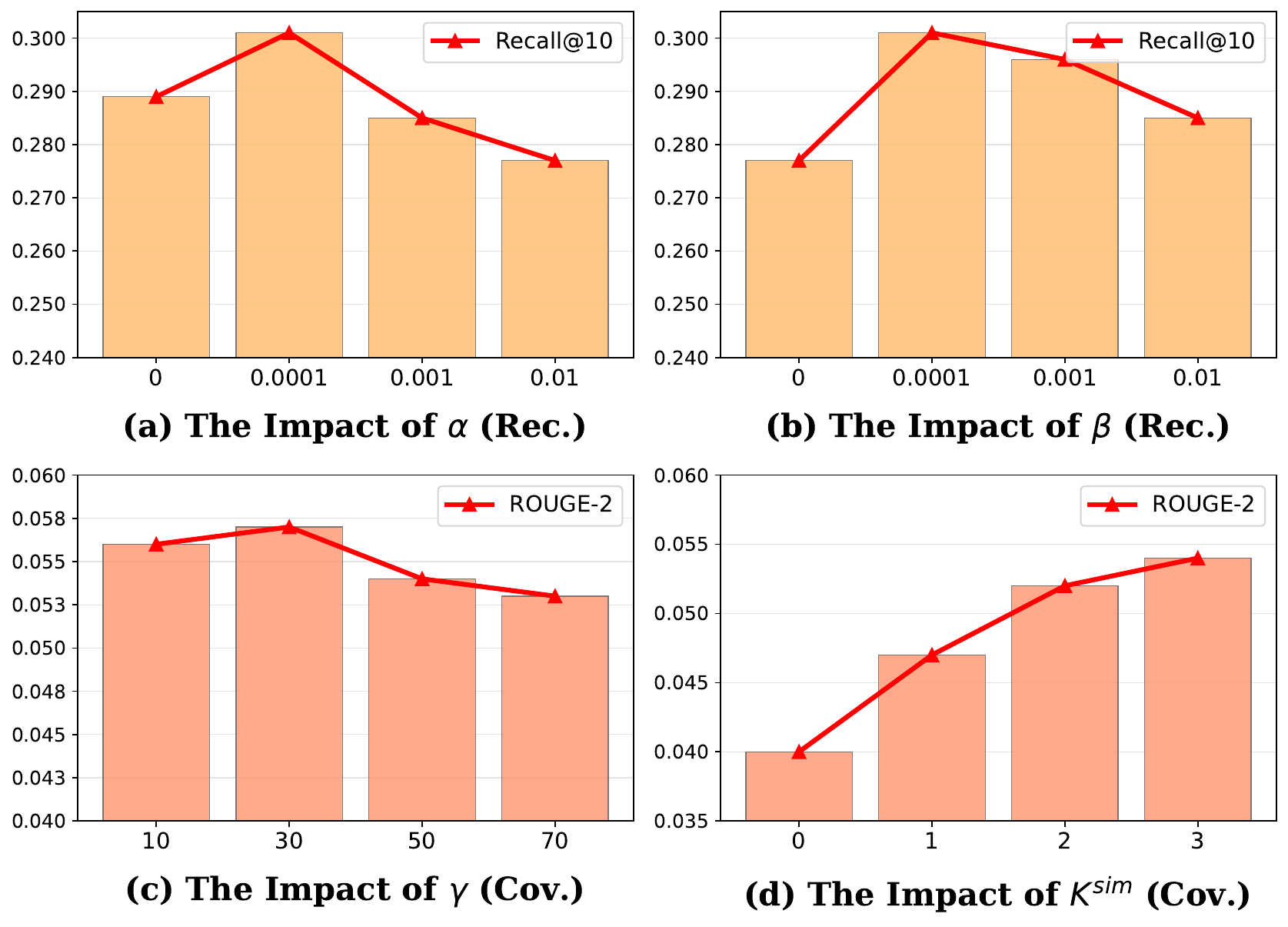}
    \caption{The impact of important hyper-parameters on INSPIRED. Similar trends can be observed on other datasets.}
    \label{fig_hyper}
\end{figure}

\subsection{Hyper-parameter Analysis}
We further analyze several key hyper-parameters of \modelname. 
For the recommendation task, we study the two contrastive coefficients used to align knowledge-based and language-based representations. 
For the conversation task, we examine the length of the visual soft prompt and the number of retrieved similar conversations.

\subsubsection{Impact of the KG-Text contrastive coefficient $\alpha$}
Figure~\ref{fig_hyper}(a) reports the results when gradually increasing the KG-Text contrastive weight $\alpha$ from $0$ to $0.01$, where $\alpha=0$ corresponds to removing the contrastive term $\mathcal{L}^{cl}_{kg\text{-}txt}$ entirely. 
As $\alpha$ increases from $0$, the recommendation performance improves, indicating that contrastive alignment between KG-based and textual representations is beneficial. 
The best performance is achieved at $0.0001$. 
However, further increasing $\alpha$ leads to performance degradation, suggesting that an overly strong contrastive objective may dominate training and hinder optimization toward the recommendation objective.

\subsubsection{Impact of the KG-Visual contrastive coefficient $\beta$}
The effect of the KG-Visual contrastive weight $\beta$ (Figure~\ref{fig_hyper}(b)) follows a similar trend: performance first increases to a peak and then drops as $\beta$ becomes larger. 
Notably, setting $\beta=0$ causes a substantial performance decrease, which suggests a larger semantic gap between visual representations and KG representations than that between text and KG. 
This observation highlights the necessity of the contrastive term for effectively aligning the visual pathway with KG semantics.

\subsubsection{Impact of the visual prompt length $\gamma$ for the current conversation}
We vary the length $\gamma$ of the visual soft prompt $\overline{\mathbf{V}}_{c}$ for the current dialogue in Figure~\ref{fig_hyper}(c). 
Overall, \modelname performs best when $\gamma=30$. 
Both overly short and overly long prompts hurt performance: a short prompt may compress the visual context too aggressively and lose salient information, whereas an excessively long prompt can introduce redundancy and distract the generator from the most useful signals.

\subsubsection{Impact of the number of retrieved conversations $K^{sim}$}
Figure~\ref{fig_hyper}(d) shows the impact of the number of retrieved similar conversations $K^{sim}$. 
When no retrieved conversation is used, response quality is noticeably limited, indicating that retrieval provides helpful complementary evidence. 
Increasing $K^{sim}$ generally improves performance, but the marginal gains diminish as $K^{sim}$ grows. 
For example, increasing $K^{sim}$ from $0$ to $1$ yields a $0.007$ improvement in ROUGE-2, whereas increasing it from $2$ to $3$ results in a gain of less than $0.003$ ROUGE-2.

\section{Conclusion}
In this paper, we propose \modelname, a screen-text-aware conversational recommender system that improves textual awareness with multi-path understanding. 
Specifically, \modelname combines fine-grained token-based understanding with vision-centric skim reading and fuses them with prompt learning for both recommendation and response generation. Extensive experiments on ReDial and INSPIRED, together with ablation and hyper-parameter analyses, demonstrate that \modelname consistently outperforms strong CRS baselines and provides a practical new direction for broaden context digestion in CRSs.

\begin{acks}
The Australian Research Council supports this work under the streams of Future Fellowship (Grant No. FT210100624), the Discovery Project (Grant No. DP240101108), and the Linkage Project (Grant No. LP230200892).
\end{acks}

\bibliographystyle{ACM-Reference-Format}
\bibliography{sample-base}

\end{document}